\documentclass{article}
\usepackage{cite}
\usepackage{graphicx}
\usepackage{dcolumn}

\begin{document}

\date{}
\title{On the interaction of a magnetic quadrupole moment with an electric field in
a rotating frame}
\author{Francisco M. Fern\'{a}ndez\thanks{%
fernande@quimica.unlp.edu.ar} \\
%EndAName
INIFTA, DQT, Sucursal 4, C. C. 16, \\
1900 La Plata, Argentina}
\maketitle

\begin{abstract}
We discuss results obtained recently for a quantum-mechanical model given by
a neutral particle with a magnetic quadrupole moment in a radial electric
field and a scalar potential proportional to the radial variable in
cylindrical coordinates that also includes the noninertial effects of a
rotating reference frame. We show that the conjectured allowed values of the
cyclotron frequency are a mere artifact of the truncation of the power
series used to solve the radial eigenvalue equation. Our analysis proves
that the analytical expression for the eigenvalues are far from correct.
\end{abstract}

\section{Introduction}

\label{sec:intro}

In a paper published recently Fonseca and Bakke\cite{FB17} discussed a
neutral particle with a magnetic quadrupole moment in a radial electric
field and a scalar potential proportional to the radial variable in
cylindrical coordinates. They also considered the noninertial effects of a
rotating reference frame. The Schr\"{o}dinger equation for this
quantum-mechanical model is separable in cylindrical coordinates and the
authors solved the radial eigenvalue equation by means of the Frobenius
(power-series) method. Upon forcing the truncation of the series they
derived exact analytical polynomial expressions for the radial
eigenfunctions as well as an exact analytical formula for the energy levels.
They concluded that there are some permitted cyclotron frequencies
determined by the angular velocity of the rotating frame, the parameter
associated to the scalar potential and the quantum numbers. In this paper we
test the validity of those results and conclusions.

In section~\ref{sec:Frobenius} we discuss the application of the Frobenius
method to the radial eigenvalue equation and in section~\ref{sec:conclusions}
we summarize the main results and draw conclusions.

\section{The Frobenius method}

\label{sec:Frobenius}

The starting point of our discussion is the eigenvalue equation
\begin{eqnarray}
&&F^{\prime \prime }(r)+\frac{1}{r}F^{\prime }(r)-\frac{l^{2}}{r^{2}}%
F(r)-r^{2}F(r)-\nu rF(r)+WF(r)=0,  \nonumber \\
&&W=\frac{4}{\alpha }\left( E+\frac{1}{2}\vartheta l+l\varpi \right)
,\;\alpha ^{2}=\vartheta ^{2}+4\varpi \vartheta ,  \nonumber \\
&&\nu =\frac{2^{5/2}a}{\sqrt{m\alpha ^{3}}},  \label{eq:dif_eq}
\end{eqnarray}
where $m$ is the mass of the particle, $E$ the energy, $l=0,\pm 1,\pm
2,\ldots $ the rotational quantum number, $a$ a constant in the scalar
potential, $\vartheta $ the cyclotron frequency and $\varpi $ the angular
velocity of the rotating frame\cite{FB17}. We can draw some straightforward
conclusions from this equation. Since the behaviour of $F(r)$ at $%
r\rightarrow 0$ and $r\rightarrow \infty $ is determined by the terms $%
l^{2}/r^{2}$ and $r^{2}$, respectively, we conclude that there are square
integrable solutions for all values of $\nu $. Therefore, there is no
restriction on the values of the cyclotron frequency $\vartheta $, contrary
to what the authors stated.

There are square integrable solutions
\begin{equation}
\int_{0}^{\infty }\left| F(r)\right| ^{2}r\,dr<\infty ,
\label{eq:square_int}
\end{equation}
for particular values of $W=W_{i,l}(\nu )$, $i=0,1,\ldots $, that are
continuous functions of $\nu $. Besides, from the Hellmann-Feynman theorem
(HFT)\cite{G32,F39}
\begin{equation}
\frac{\partial W}{\partial \nu }=\left\langle r\right\rangle >0,
\label{eq:HFT}
\end{equation}
we conclude that each $W_{i,l}(\nu )$ is an increasing function of $\nu $.

Fonseca and Bakke\cite{FB17} focused on polynomial solutions to equation (%
\ref{eq:dif_eq}) that we discuss in what follows. If we look for a solution
of the form
\begin{equation}
F(r)=r^{s}\exp \left( -\frac{r^{2}}{2}-\frac{\nu r}{2}\right)
\sum_{j=0}^{\infty }c_{j}r^{j},\;s=|l|,  \label{eq:F(r)_series}
\end{equation}
we conclude that the expansion coefficients should satisfy the three-term
recurrence relation
\begin{eqnarray}
c_{j+2} &=&A_{j}c_{j+1}+B_{j}c_{j},\;j=-1,0,1,\ldots ,\;c_{-1}=0,  \nonumber
\\
A_{j} &=&\frac{\nu \left( 2j+2s+3\right) }{2\left( j+2\right) \left[
j+2\left( s+1\right) \right] },  \nonumber \\
B_{j} &=&-\frac{4W-8j+\nu ^{2}-8\left( s+1\right) }{4\left( j+2\right)
\left[ j+2\left( s+1\right) \right] }.  \label{eq:TTRR}
\end{eqnarray}
In order to obtain a polynomial solution of order $n$, $n=0,1,\ldots $, we
require that $c_{n}\neq 0$, $c_{n+1}=0$ and $c_{n+2}=0$ that leads to $%
B_{n}=0$. From the last equality we obtain
\begin{equation}
W=W_{l}^{(n)}=2(n+s+1)-\frac{\nu ^{2}}{4},  \label{eq:W_l^(n)}
\end{equation}
that leads to
\begin{equation}
B_{j}=B_{j,n}=\frac{2\left( j-n\right) }{\left( j+2\right) \left[ j+2\left(
s+1\right) \right] }.  \label{eq:B_(j,n)}
\end{equation}
We immediately realize that something is amiss because the eigenvalues $%
W_{l}^{(n)}$ do not satisfy the HFT (\ref{eq:HFT}).

Since $c_{n+1}$ is a polynomial function of $\nu $ of degree $n+1$, the
second condition $c_{n+1}=0$ leads to $n+1$ particular values of $\nu $, $%
\nu _{n,i.l}$, $i=1,2,\ldots ,n+1$. From these roots Fonseca and Bakke\cite
{FB17} concluded that ``only specific values of the cyclotron frequency $%
\vartheta $ are permitted''. However, they did not attempt to investigate
the actual meaning of these values of the model parameter $\nu $. For
convenience, here we organize them in decreasing order $\nu _{n,i,l}>\nu
_{n,i+1,l}$. We appreciate that for each value of $n$ the resulting
eigenvalues (\ref{eq:W_l^(n)})
\begin{equation}
W_{l}^{(n,i)}=2(n+s+1)-\frac{\nu _{n,i,l}^{2}}{4},  \label{eq:W_l^(n,i)}
\end{equation}
are located on an inverted parabola. When $n+1$ is odd there is a root $\nu
=0$ that leads to the exact eigenvalues of the harmonic oscillator.

The truncation of the series in equation (\ref{eq:F(r)_series}) leads to
particular polynomial solutions of the form
\begin{equation}
F_{l}^{(n,i)}(r)=r^{|l|}\exp \left( -\frac{r^{2}}{2}-\frac{\nu _{n,i,l}r}{2}%
\right) \sum_{j=0}^{n}c_{j,n,l}r^{j}.  \label{eq:F(r)_poly}
\end{equation}
They are square integrable but there are other solutions that satisfy the
condition (\ref{eq:square_int}) that do not have polynomial factors. Since
Fonseca and Bakke overlooked the latter, they drew the wrong conclusion
mentioned above.

This kind of problems is commonly called quasi-exactly solvable or
conditionally solvable because one obtains eigenvalues $W$ only for
particular values of $\nu $. They have been studied by several authors (see,
for example, the review by Turbiner\cite{T16} and the references therein).
Fonseca and Bakke\cite{FB17} seemed to be unaware of this fact and appeared
to believe that the only quadratically integrable solutions to equation (\ref
{eq:dif_eq}) are the polynomial ones (\ref{eq:F(r)_poly}). For this reason
they concluded, wrongly, that there are allowed values of the cyclotron
frequency $\vartheta $. The true fact is that the allowed energies $E_{n,l}$
are continuous functions of this model parameter.

There is no doubt that $W_{l}^{(n,i)}$ and $F_{l}^{(n,i)}(r)$ are
eigenvalues and eigenfunctions, respectively, of the differential equation (%
\ref{eq:dif_eq}). A question now arises about the connection between the
eigenvalues $W_{l}^{(n,i)}$ of such polynomial solutions and the actual
eigenvalues $W_{j,l}(\nu )$ mentioned above. Taking into account the HFT (%
\ref{eq:HFT}) and the convenient ordering of the roots $\nu _{n,i,l}$ chosen
above, we conclude that $W_{l}^{(n,i)}\left( \nu _{n,i,l}\right)
=W_{i-1,l}\left( \nu _{n,i,l}\right) $; in other words, $W_{l}^{(n,i)}\left(
\nu _{n,i,l}\right) $ is a particular point on the continuous curve $%
W_{i-1,l}\left( \nu \right) $.

\begin{figure}[tbp]
\begin{center}
\includegraphics[width=9cm]{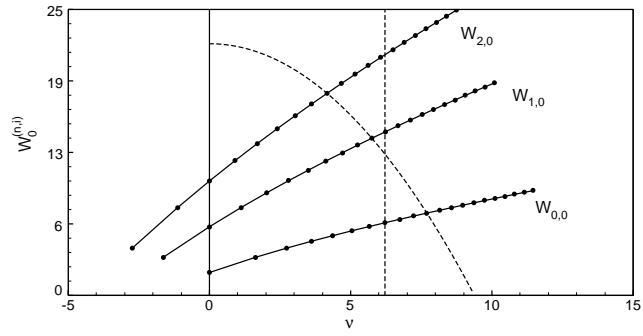}
\end{center}
\caption{Eigenvalues $W_0^{(n,i)}$ from the truncation method and actual
eigenvalues $W_{n,0}(\nu)$ of the differential equation (\ref{eq:dif_eq}) }
\label{fig:Wn0}
\end{figure}

Figure~\ref{fig:Wn0} shows some selected points $W_{0}^{(n,i)}\left( \nu
_{n,i,0}\right) $, $n\leq 22$, $i\leq \min (n+1,3)$, connected by continuous
lines that draw the curves $W_{j,0}(\nu )$, $j=0,1,2$. The inverted parabola
$W_{0}^{(10)}(\nu )=22-\nu ^{2}/4$ connects some of the solutions $%
W_{0}^{(10,i)}$ given by equation (\ref{eq:W_l^(n,i)}). The intersections of
the vertical dashed line with the curves $W_{j,0}(\nu )$ are the actual
eigenvalues of the quantum model with a given value of the parameter $\nu $.
Such a vertical line passes through, at most, one value of $%
W_{0}^{(n,i)}\left( \nu _{n,i,0}\right) $ as shown in figure~\ref{fig:Wn0}.

In order to obtain $W_{j-,l}(\nu )$ for $\nu _{n,j,l}<\nu <\nu _{n+1,j,l}$,
we simply resort to any suitable interpolation method. For example, from
least squares we obtain
\begin{eqnarray}
W_{0,0}(\nu ) &=&2+0.8523002844\nu -0.02975046592\nu ^{2}+0.0008706577439\nu
^{3},  \nonumber \\
W_{1,0}(\nu ) &=&6+1.547791990\nu -0.04202730246\nu ^{2}+0.001218822726\nu
^{3},  \nonumber \\
W_{2,0}(\nu ) &=&10+2.010156364\nu -0.04562156939\nu ^{2}+0.001269456909\nu
^{3},  \nonumber \\
&&  \label{eq:W_(n,0)_fit}
\end{eqnarray}
that are sufficiently accurate in the range of $\nu $ values shown in figure~%
\ref{fig:Wn0}.

\section{Conclusions}

\label{sec:conclusions}

The eigenvalues $W_{i,l}(\nu )$ of the differential equation (\ref{eq:dif_eq}%
) are continuous functions of the model parameter $\nu $. Consequently, the
energy eigenvalues $E_{n,l}$ are continuous functions of the cyclotron
frequency $\vartheta $ that can take any physically acceptable value. The
truncation of the series in equation (\ref{eq:F(r)_series}) only yields
eigenvalues $W_{l}^{(n.i)}$ for particular values $\nu _{n,i,l}$. From this
particular values of $\nu $ Fonseca and Bakke\cite{FB17} concluded, wrongly,
that there are allowed or permitted values of the cyclotron frequency $%
\vartheta $. The eigenvalues $\mathcal{E}_{n,l}$ in equation (17) of the
paper by Fonseca and Bakke are meaningless because the model parameters
change with the quantum numbers $n$ and $l$. More precisely, $\mathcal{E}%
_{n,l}$ and $\mathcal{E}_{n^{\prime },l^{\prime }}$ are energy eigenvalues
of different physical problems.

\end{document}